\documentclass[preprint2]{aastex}
\usepackage{amsmath}
\newcommand{\be}{\begin{equation}}
\newcommand{\ee}{\end{equation}}
\begin{document}

\title{Lossy compression of weak lensing data}
\author{R.~Ali Vanderveld\altaffilmark{1,2,3}, 
Gary M.~Bernstein\altaffilmark{4}, 
Chris Stoughton\altaffilmark{5},
Jason Rhodes\altaffilmark{2,3}, 
Richard Massey\altaffilmark{6}, 
David Johnston\altaffilmark{5}, and
Benjamin M. Dobke\altaffilmark{2,3}}

\altaffiltext{1}{Kavli Institute for Cosmological Physics, Enrico Fermi Institute, University of Chicago, Chicago, IL 60637}
\altaffiltext{2}{Jet Propulsion Laboratory, California Institute of Technology, Pasadena, CA 91109}
\altaffiltext{3}{California Institute of Technology, 1200 East California Boulevard, Pasadena, CA 91125}
\email{rav@kicp.uchicago.edu}
\altaffiltext{4}{Dept. of Physics and Astronomy, University of Pennsylvania, Philadelphia, PA 19104}
\altaffiltext{5}{Center for Particle Astrophysics, Fermi National Accelerator Laboratory, PO Box 500, Batavia, IL 60510}
\altaffiltext{6}{Institute for Astronomy, Royal Observatory, Blackford Hill, Edinburgh EH9 3HJ, UK}

\begin{abstract}

Future orbiting observatories will survey large areas of sky in order to constrain the physics of dark matter and dark energy using weak gravitational lensing and other methods.  Lossy compression of the resultant data will improve the cost and feasibility of transmitting the images through the space communication network. We evaluate the consequences of the lossy compression algorithm of Bernstein et al.~(2010) for the high-precision measurement of weak-lensing galaxy ellipticities. This square-root algorithm compresses each pixel independently, and the information discarded is by construction less than the Poisson error from photon shot noise. For simulated space-based images (without cosmic rays) digitized to the typical 16 bits per pixel, application of the lossy compression followed by image-wise lossless compression yields images with only $2.4$ bits per pixel, a factor of 6.7 compression.  We demonstrate that this compression introduces no bias in the sky background.  The compression introduces a small amount of additional digitization noise to the images, and we demonstrate a corresponding small increase in ellipticity measurement noise. The ellipticity measurement method is biased by the addition of noise, so the additional digitization noise is expected to induce a multiplicative bias on the galaxies' measured ellipticities.  After correcting for this known noise-induced bias, we find a residual multiplicative ellipticity bias of $m\approx -4\times 10^{-4}$.  This bias is small when compared to the many other issues that precision weak lensing surveys must confront, and furthermore we expect it to be reduced further with better calibration of ellipticity measurement methods.

\end{abstract}


\section{Introduction}

Weak gravitational lensing, whereby we measure how the images of field galaxies are distorted by the intervening matter distribution, is a powerful tool for probing the physics of the ``dark sector" \citep{DETF,SWG}, with very promising results for large-scale cosmology in recent years, e.g.~\cite{COSMOS1,COSMOS2,CFHTLS1,CFHTLS2,COSMOS3}.  As such, this technique is expected to be at the forefront of efforts to constrain the nature of dark matter and dark energy, and the most powerful experiments will utilize space observatories conducting surveys over the largest possible area of sky \citep{AR}. The Wide-Field Infrared Survey Telescope (WFIRST)\footnote{\tt http://wfirst.gsfc.nasa.gov/} and Euclid\footnote{\tt http://sci.esa.int/science-e/www/area/index.cfm?fareaid=102} are proposals for such large-area space experiments. 

Data compression has many benefits. It allows a reduction in onboard storage requirements, which lowers cost and lowers power requirements and heat output, thereby making the mission design simpler.  It also lowers the need for downlink time which is expensive on the Deep Space Network (DSN).  For example, for a WFIRST weak lensing survey taking data every 180 seconds with 36 detectors each comprised of $2048\times 2048$ pixels with an uncompressed 16 bits per pixel, we would need to downlink 135 GB of imaging data (plus spectra and calibration data) per day with the DSN's data rate of $150~{\rm MB}/{\rm second}$. The full range of benefits of data compression are complex and depend on mission design, but certainly the compression option allows flexibility in that design. The drawback of compression is possible loss of crucial information, which is what we explore in this study. 

Note that CCD data already suffer some lossy ``compression" when the analog voltage representing the accumulated photon count is digitized into Analog-to-Digital Units (ADUs) for storage and transmission. One of the more popular schemes for additional lossy compression is called ``square-root" compression \citep{GS}, which, as the name implies, takes the square root of the pixel ADU values and truncates them so that they can be represented by fewer bits per pixel. Square-root compression is attractive because the additional error introduced by truncation is a fixed fraction of the Poisson noise already present in the photoelectron signal. Our goal in this study is to find how the application of this square-root compression algorithm modifies weak lensing data and the inferences that we would draw from them, in the absence of any attempts to correct for the effects.  We do this using simulated sky images created with a ``shapelets"-based pipeline \citep{Sim1,Sim2,Sim3}.  We apply the square-root compression scheme of \citet{Gary}, and build upon that work by answering two questions: (1) Does this compression scheme bias the sky background? (2) Knowing the background, is shape information conserved?

The compression algorithm essentially re-bins the pixel values more coarsely than the original digitization. For weak lensing surveys we are interested mainly in faint objects, so the effect of lossy compression on both the transmission rate and the image fidelity should be primarily determined by the coarseness of this re-binning at the sky background level. The critical parameter is
\be
b\equiv\frac{\sigma_{\rm sky}}{N_{\rm step}}
\label{Bernstein}
\ee
where $\sigma_{\rm sky}$ is the RMS of sky pixels in the image and $N_{\rm step}$ is the number of input ADU values that are encoded to a common output value by the compression algorithm in the vicinity of the sky level.  The ratio of these quantities is the number of bits that span the sky noise in the compressed image.  The higher this number, the better we expect the image properties to be reproduced in the decompressed version.  In particular we expect poor results when $b<1$.

For a next-generation weak lensing experiment, the cosmological biases caused by a multiplicative bias $m$ in measured galaxy ellipticities will be safely below the experiment's statistical errors if $m<10^{-3}$ \citep{AR}.  We simulate a large enough sample of galaxies to probe this bias requirement, and our goal is to find if total (lossy plus lossless) compression by a factor of $\sim 3$ can be attained without violating it.  We find that lossy compression at $b=1$ more than satisfies the compression requirement, does not bias the sky background, and induces an RMS shift in galaxy shape of only 0.027, completely negligible when added in quadrature the intrinsic ellipticity spread of roughly $0.3$ that sets a floor on weak lensing measurements.

On the other hand, we find that the data compression/decompression (codec) procedure biases the magnitude of measured ellipticities, thereby inducing a multiplicative bias on the apparent weak lensing shear. The RRG ellipticity measurement method we use \citep{RRG} is known to be biased by the addition of noise, and thus we do expect the digitization noise inherent to the compression to induce a multiplicative bias on the galaxies' measured ellipticities.  When the codec's multiplicative bias is corrected for this known shortcoming of the RRG method, we find an excess compression-induced multiplicative ellipticity bias of $m\approx -4\times 10^{-4}$ for $b=1$, thereby meeting the requirement $|m|<10^{-3}$ by a factor that we expect to be increased with appropriate calibration, as discussed later in the paper.

This paper is organized as follows.  In Section 2 we discuss our study, including a basic review of the lossy compression scheme we use, our test images, and our weak lensing analysis pipeline.  In Section 3 we give our results, and in Section 4 we provide a discussion and recommendations. All quoted errors and plotted error bars correspond to one standard deviation for the entirety of the paper.

\section{Method}

\subsection{Compression scheme}

We use the compression scheme, including bias correction, as described in \citet{Gary}.  We provide a brief description here.  We assume that the telescope design has readout performed by electronics to produce one $16$ bit number per pixel.  Computing on board will reduce this to fewer bits per pixel using a lossy compression algorithm.  Further computing will apply a lossless compression algorithm, and the goal is that we can achieve an overall compression factor (from the original $16$ bits per pixel) of $\sim 3$. The lossy compression step can be expressed as a lookup table, as the mapping for a given pixel value is always the same. Note also that we must apply the lossy compression before the lossless, as the lossy works on each pixel independently and the lossless step would interfere with this mapping.

The square-root algorithm for lossy data compression described in \citet{GS} transforms an input value $x$ to a compressed value $y$ as
\be
y = {\rm int}\left(0.5+A+\sqrt{B*x-C}\right)~,
\label{comp}
\ee
where $A$, $B$, and $C$ are constants specified by the maximum and minimum values of the input and compressed values, and the ${\rm int}$ function rounds to the nearest integer.  The compression transformation applies Eq.~(\ref{comp}) with the appropriate values of $A$, $B$, and $C$ to calculate the compressed value. The decompression algorithm returns the average of all uncompressed values that yield the compressed value determined by Eq.~(\ref{comp}).  The lossy compression does not reproduce the input parameters exactly, by definition.  The codec process has a similar effect on the data as do read noise in the readout electronics or Poisson statistics.

Bernstein et al.~(2010) refine the basic square-root codec in Eq.~(\ref{comp}) with: choices for A, B, and C which maintain constant $\sigma / N_{\rm step}$ at any signal level for given detector gain and read noise; a prescription for slight departures from (\ref{comp}) to produce a codec that has uniform behavior of $N_{\rm step}$ as the signal increases; and a correction to the decompressed values which eliminates small biases in the mean signal introduced by the codec process. We will primarily focus on an implementation of the square-root compression algorithm that yields $b=1$, which we naively expect to provide the best compromise between our desires for a high compression level but for low image degradation, but we will also do some tests with a coarser $b=0.71$ and a finer $b=1.41$ level of compression. The lossy compression algorithms used in this paper can all be implemented as simple lookup tables following gain $g$ and digitization of the analog detector output.  Using the notation of Bernstein et al.~(2010), these three codecs are constructed as follows. Output code $i$ is assigned to all input integers in the range  $N_i \pm (\Delta_i-1)/2$, with $\Delta_i$ an integer giving $N_{\rm step}$ for this output code.  If we define the range step $\Delta^\prime_i = \Delta_{i+1}-\Delta_i$, code $i$ is decoded to the (half-)integer value $N_i$, plus a small correction $\approx \Delta^\prime_i/6$ that eliminates a small reconstruction bias.  Most of the results in this paper will use a codec with $g=0.5$ electrons per ADU and $\Delta^\prime_i=1$, which yields a codec with minimal reconstruction bias, a code step $N_{\rm step}=\sigma$ and $b=1$, very similar to a choice of $B=2$ in Eq.~(\ref{comp}).  We will also at times employ two other codec schemes: (1) $b=1.41$, for which $\Delta^\prime_i$ follow the sequence $\{0,1,0,1,0,1,\ldots\}$; (2) $b=0.71$, which has the same lookup table as the original codec but with $g=1$ electron per ADU before digitization.  See Bernstein et al.~(2010) for a complete description of the compression algorithm and the exact correction factors for decompression.

For a representative $2000\times 2000$ pixel test image used in our study, in the absence of cosmic rays, the readily-available lossless compression scheme {\tt bzip2}\footnote{\tt http://bzip.org/} alone reduces the file size from the original $16.0$ MB (or equivalently, $32$ bits per pixel) to $3.2$ MB ($6.4$ bits per pixel), a reduction by a factor of $5$ and a compression level which depends critically on the gain and sky and noise levels, specified in the next subsection. On the other hand, lossy compression alone reduces the file size to $8.0$ MB ($16$ bits per pixel), a factor of $2$ reduction. The combination of lossy compression followed by {\tt bzip2} reduces the file size to $1.4$ MB ($2.8$ bits per pixel), $1.2$ MB ($2.4$ bits per pixel), and $1.0$ MB ($2.0$ bits per pixel) for $b=1.41$, $1$, and $0.71$, respectively. This is similar to the theoretically expected optimum value for Gaussian-noise images, as per Bernstein et al.~(2010). Moreover, also as noted in Bernstein et al.~(2010), {\tt bzip2} is not very robust, in that a single-bit transmission error can lead to loss of a full image; a better algorithm, used on over 25 space missions, is CCSDS 121B \citep{CCSDS}. Bernstein et al.~(2010) found CCSDS 121B to yield very similar filesizes for weak lensing images, to within $0.1$ bits per pixel of the {\tt bzip2} results.

\begin{figure}
\plotone{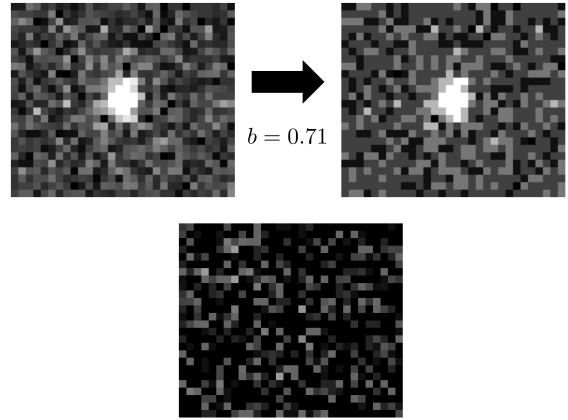}
\caption[]{A patch of an image, shown on the top left before compression and shown on the top right after codec with $b=0.71$. The residuals (multiplied by a factor of $5$) are shown on the bottom.} 
\label{byeye} 
\end{figure}

In Figure~\ref{byeye}, we show an example of a patch of an image that includes an object before compression on the top left, and we show the same patch after the aforementioned codec scheme with $b=0.71$ on the top right, with the residuals (multiplied by a factor of $5$ for clarity) on the bottom.  The coarser greyscale is apparent even by eye in the background noise from this rather extreme compression level, as one can see the smaller number of grey levels in use. This re-binning is less severe for the higher values of $b$ that we use in the remainder of this paper. 

\begin{figure}
\plotone{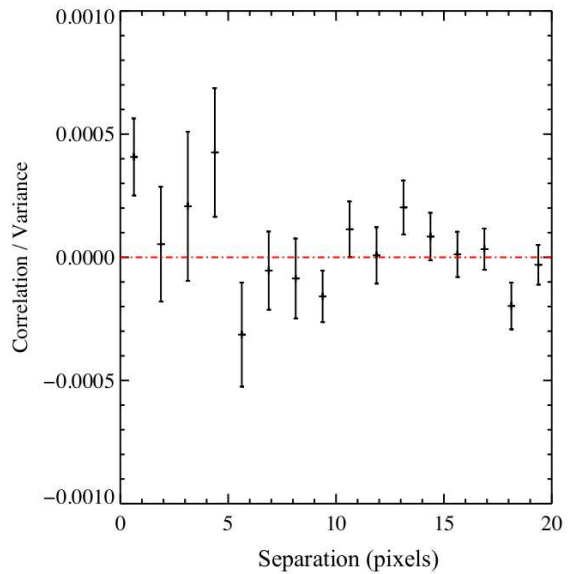}
\caption[]{The ratio of the two-point correlation function to the variance (the zero lag correlation function), plotted as a function of distance in pixels, for the difference between an original image and its codec counterpart for $b=1$.}
\label{Dave} 
\end{figure}

This lossy compression algorithm is designed to remove bits per pixel which are shot noise, which is equivalent to adding on small amount of extra noise. Therefore the resulting compressed images should be comparable to images with a slightly lower exposure time, the penalty being a factor of $1+b^{2}/12$, which is $8\%$ if $b=1$. The compression is done independently for each pixel, so naively one would expect this added noise to be white.  We provide evidence of this in Figure~\ref{Dave}, which is a plot of the ratio of the two-point correlation function to the variance (i.e.~the zero lag correlation function), as a function of distance in pixels, for the difference between an original image and its codec counterpart for $b=1$.  As we can see, the correlations are all at least a factor of $10^{3}$ smaller than the variance, which is consistent with the properties of white noise. Further note that the residuals in Figure~\ref{byeye} appear consistent with white noise.

\subsection{Images}

To test whether our codec algorithm biases the sky background, we make images that are purely Poissonian sky noise plus read noise.  We then run these images through the aforementioned codec scheme, plus de-biasing, and compare the mean of the codec image to the mean of the original. The images that we will use for the galaxy shape portion of this study are simulated with the shapelets method, described in \citet{Sim3} and used in the Shear TEsting Program (STEP) collaboration shear extraction tests \citep{STEP2} and in \citet{High}.  These images are randomly generated, based on Hubble Ultra Deep Field (UDF) data.  Survey characteristics such as mirror size, exposure time, pixel scale, galaxy and star number density, Point Spread Function (PSF) type, and noise are freely specifiable by the user.  A known external shear can also be added to each image.

Following the STEP methodology, we have manufactured a large set of space-like images meant to be similar to the data set resulting from a survey like WFIRST or Euclid.  We use the same bandpass as the COSMOS HST/ACS survey data so that we can use the same model for the expected galaxy population.  We also use $0.07$ arcsecond pixels, an 800 second exposure time, a PSF with a $50\%$ encircled-energy radius of $0.15$ arcseconds, and an effective imager collecting area of $0.83~{\rm m}^{2}$.  We assume a sky and dark current background of 45 electrons plus Poisson noise and a read noise of 4 electrons.  We don't put any shear into the images, as the goal here is not to extract a shear signal but instead to see how the codec procedure changes raw galaxy shapes.

Throughout this paper, we will refer to any unaltered images as the ``original" images.

\subsection{Weak lensing pipeline}

All of the original galaxy images are run through the codec algorithm described in Section 2.  Then the original and codec images are run through the following weak lensing pipeline:

\begin{itemize}

\item {\tt SExtractor} \citep{sextractor} is run only on the original, uncompressed, images.  The resulting detections and sky backgrounds are then used for the weak lensing analyses of both the original and the codec images.  In other words, we do not run {\tt SExtractor} on the codec images and we instead use the {\tt SExtractor} catalogs produced from the original images on everything. This ensures that consistent object lists and sky levels are used for the codec and no-codec images.

\item Galaxy shapes are measured in both original and codec images with the RRG method \citep{RRG}.

\item Size, ellipticity, and S/N cuts are done from both the original and codec images.  Any objects that are cut in that stage in either its original or codec form are not included in our analysis.  We cut all galaxies with a S/N less than $10$, a size less than $1.25$ times the PSF size, or nonphysical (i.e.~greater than $1$) ellipticities.

\end{itemize}

RRG is based on the KSB$+$ shape measurement method \citep{Kaiser, Hoekstra} which measures Gaussian-weighted multipole image moments,
\be
J_{ij}=\int d^2\theta w\left(\theta\right)I\left(\theta\right)\theta_i\theta_j~,
\ee
where $w$ is a Gaussian weighting function and $\theta$ is chosen such that the weighted barycenter is zero.  The resulting ellipticity is
\be
\left(e_1,e_2\right)=\frac{1}{J_{xx}+J_{yy}}\left(J_{xx}-J_{yy},2J_{xy}\right)~
\ee
and we define the size to be
\be
d=\sqrt{\frac{1}{2}\left(J_{xx}+J_{yy}\right)}~.
\label{size}
\ee
We do not perform PSF deconvolution because we are looking only at the  shape change induced by the compression process.  PSF deconvolution can induce biases larger than the effects we are trying to measure here (see, e.g.~the results of the GREAT08 challenge in \citet{GREAT08}). We measure only the raw shape as parameterized by the two component ellipticity defined above and determine how this is affected by the codec process.

\section{Results}

\subsection{Sky background}

\begin{figure}
\plotone{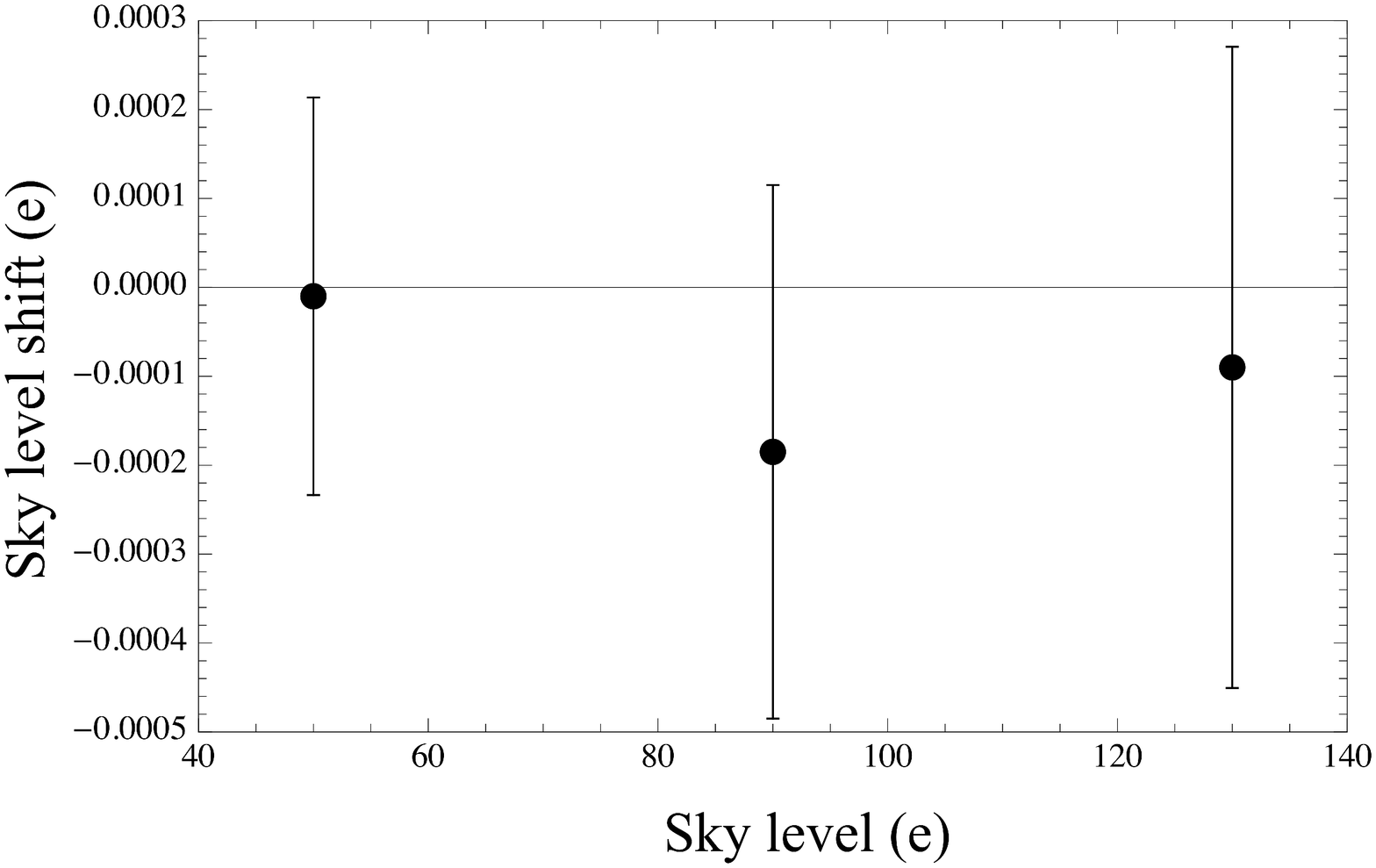}
\caption[]{The mean shift in the sky level (in units of electrons) due to codec with $b=1$, plotted as a function of the sky level (also in electrons) in the uncompressed images. Each data point corresponds to $10^9$ pixels.}
\label{skybias} 
\end{figure}

Applying this lossy compression scheme to astronomical images re-bins the sky background.  Does this process bias the measured sky level?  As was found in \citet{Gary}, a codec with equally-spaced steps should not, and for other codec schemes it is possible to de-bias during the reconstruction process.  Using the procedure as described in Section 2, with $10^9$ pixels and using our codec with $b=1$, we find the sky background to be amplified by a factor of $(2\pm 3)\times 10^{-6}$ for the fiducial survey we consider here.  This is negligible.  We find similarly insignificant biases when trying other sky background levels, as can be seen in Figure~\ref{skybias}.

This test also shows that whenever we have many pixels with the same underlying value but different noise, the bias in the mean of them is small, below the shot-noise level. Thus, with $N$ copies of a galaxy image, each with independent noise realizations, the difference between a stacked codec image and the original image falls as expected.

\subsection{Galaxy shapes}

\begin{figure}
\plotone{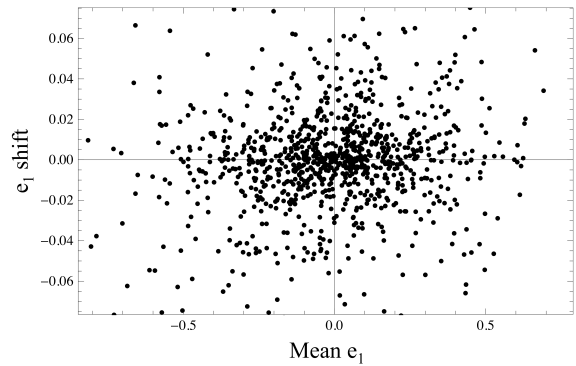}
\caption[]{Scatter plot of shifts in $e_1$ resulting from codec with $b=1$, plotted as a function of the mean $e_1$ before and after codec.}
\label{scatterplot} 
\end{figure}

Given perfect knowledge of the sky background, how are galaxy shapes affected by this codec procedure?  We probe this question with $\sim 2.5\times 10^{6}$ simulated galaxies for our fiducial survey and the weak lensing pipeline described above. Figure~\ref{scatterplot} is a scatter plot of resulting $e_1$ shifts as a function of the mean $e_1$ before and after codec, for $b=1$ and a representative subsample of $1000$ galaxies.

\begin{figure}
\plotone{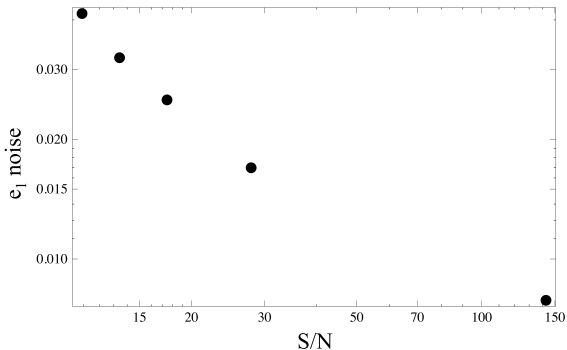}
\caption[]{Standard deviation of $e_1$ shifts resulting from codec with $b=1$, as a function of galaxy S/N.}
\label{e1noise} 
\end{figure}

Firstly, we find a negligible added shape noise. Such added noise would decrease the statistical power of the survey but would not add a bias. The additional noise on the ellipticities due to codec digitization with $b=1$ is, by design, a factor of $\sqrt{12}$ lower than the noise from photon statistics and read noise \citep{Gary}, which is in turn typically lower than the intrinsic shape noise. For this compression level, the standard deviation of the ellipticity shifts induced by the codec is $0.027$. Such added noise is an order of magnitude smaller than the ellipticity spread due to intrinsic shape noise, and it depends on galaxy S/N as can be seen in Figure~\ref{e1noise} for $e_1$. We find similar results for $e_2$. 

\begin{figure}
\plotone{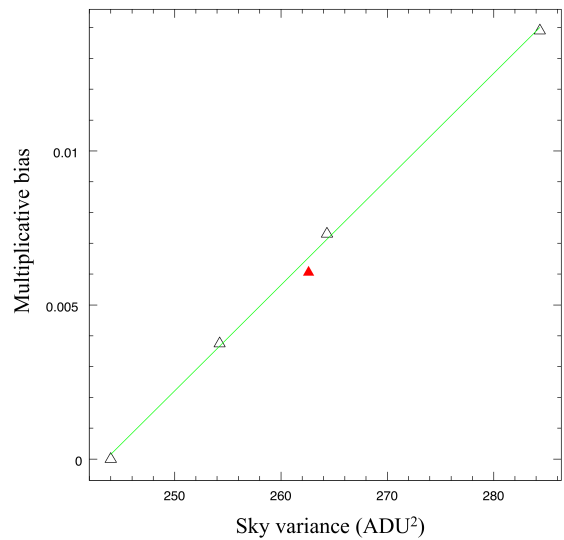}
\caption[]{The ellipticity multiplicative bias as a function of variance for our simulations. The open triangles correspond to the results for our noise-equalized images (no codec) and the green solid line is the result of fitting the open triangle points to a straight line. The red filled triangle corresponds to the codec image result for $b=1$.}
\label{dmdv} 
\end{figure}

We also look for offset and bias, as in the STEP papers \citep{STEP1,STEP2}. When $\sigma=N_{\rm step}$, the discreteness of the codec will add an additional variance of approximately $1/(12b^2)$ of the original image's noise variance \citep{Gary}. To test the effects of adding this slight increase in noise level, in the absence of compression, we create ``noise-equalized" images by adding this level of additional Gaussian random noise to the original images. We then measure the offset and bias that results from this noise addition as follows. For a given galaxy, let $e_{i}^{{\rm o}}$ be its ellipticity as measured in the original images and $e_{i}^{{\rm f}}$ be what we measure from these noise-equalized images, where $i=1,2$.  We then fit the difference as a function of the mean:
\be
e_1^{{\rm f}}-e_1^{{\rm o}}=m_1\left(\frac{e_1^{{\rm f}}+e_1^{{\rm o}}}{2}\right)+c_1
\label{e1}
\ee
\be
e_2^{{\rm f}}-e_2^{{\rm o}}=m_2\left(\frac{e_2^{{\rm f}}+e_2^{{\rm o}}}{2}\right)+c_2~.
\label{e2}
\ee
Note that we fit as a function of the mean (as opposed to as a function of the original) so as to symmetrize the equations and avoid additional biases that come from the regression of two noisy variables. We find the biases resulting from this added noise for $b=1.41$, $1$, and $0.71$. In other words, for each of these values of $b$, we add the expected appropriate amount of excess noise, measure galaxy shapes before and after, and perform the fits in Eqs.~(\ref{e1}) and (\ref{e2}) above. For all $2.5\times 10^6$ of the galaxies lumped together into one large sample, we use standard chi-squared linear regression to find that all offsets are consistent with (i.e.~within $1-2\sigma$ of) zero. However, we find non-negligible multiplicative biases, with $m_1=m_2$ to within our statistical uncertainties. We plot these biases as a function of total sky variance as the open triangles in Figure~\ref{dmdv}. The error bars are smaller than the symbols, and we include the zero added noise data point. Fitting these data to a line, we find
\be
m = \alpha + \beta \left(\frac{v}{{\rm ADU}^2}\right)~,
\label{mfit}
\ee
where $\alpha=-0.084 \pm 0.002$, $\beta=0.000344 \pm 6\times 10^{-6}$, and $v$ is the sky variance in ${\rm ADU}^2$. This fit is plotted as the solid (green) line in Figure~\ref{dmdv}. Note that this relation is specific to the shape measurement pipeline used here.

We now have a relation for the multiplicative bias as a function of image variance, Eq.~(\ref{mfit}), found by adding excess noise in the absence of any compression. Now we need to check how well the above theoretical noise estimates correspond to the noise level actually seen resulting from our codec procedure. We do this with blank-sky images, which have a variance of $244~{\rm ADU}^2$ in our simulations in the absence of added noise or codec. From averaging over $3\times 10^6$ pixels of blank sky, we find the codec images to have a variance of $262.5\pm 0.2~{\rm ADU}^2$ for $b=1$, which is $1.007 \pm 0.001$ lower than naively predicted above. We find a similar deficit by a factor of $1.005$ for $b=1.41$. Note that some mismatch is to be expected, as this added variance was estimated while assuming that the digitization error is uniformly distributed between $+1/2$ and $-1/2$ the width of the code step, whereas this is not quite true since the noise distribution is not flat. Furthermore, the digitization noise from the codec is not uniform because the input data are already digitized, so the induced errors are only a few possible integer values.

Plugging our result for the codec-induced variance for $b=1$, $v=262.5\pm 0.2~{\rm ADU}^2$, into Eq.~(\ref{mfit}), we predict a multiplicative bias of $0.0065 \pm 0.0001$. We then measure this bias by fitting Eqs.~(\ref{e1}) and (\ref{e2}) to a line for our $2.5\times 10^6$ simulated galaxies, where now superscript ``f" denotes ellipticities measured from the codec images and once again ``o" denotes ellipticities in the original unaltered images. We find both offsets to be within $1\sigma$ of zero, and $m_1=0.00605\pm 0.00007$ and $m_2=0.00607\pm 0.00007$. These measurements are represented in Figure~\ref{dmdv} as the filled red triangle. Hence, after correcting for the added noise, we find a residual multiplicative bias of $-0.0004 \pm 0.0001$ for the $b=1$ case. Performing a similar analysis for the finer compression scheme with $b=1.41$, we similarly find no statistically-significant offsets, and $m_1=0.00363 \pm 0.00005$ and $m_2=0.00360 \pm 0.00005$; from Eq.~(\ref{mfit}) we would have expected $m=0.0032 \pm 0.0001$ for this case. Thus we find that, after correcting for the known bias due to the additional digitization variance, codec with $b=1.41$ induces an excess multiplicative bias of $0.0004 \pm 0.0001$. 

\begin{figure}
\plotone{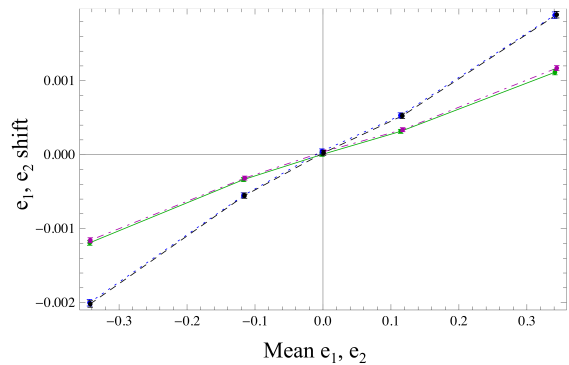}
\caption[]{Shifts in $e_1$ (blue, dotted line) and $e_2$ (black, dashed line) for codec with $b=1$ as a function of the mean $e_1$ and $e_2$, respectively, when the galaxies are sorted into five wide bins and the shifts (codec vs.~original) are averaged. For comparison, the same is plotted for $e_1$ (green, solid line) and $e_2$ (magenta, dot-dashed line) for the less-severe codec with $b=1.41$.}
\label{e1e2shifts} 
\end{figure}

\begin{figure}
\plotone{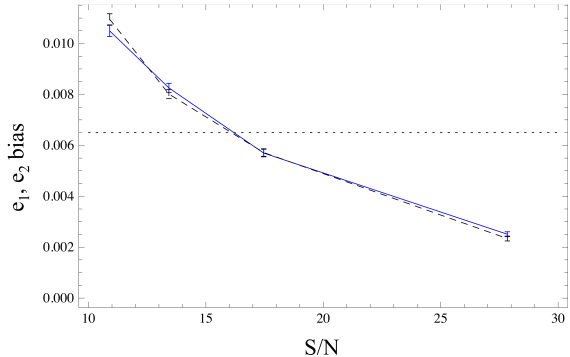}
\caption[]{Multiplicative bias for $e_1$ (blue, solid line) and $e_2$ (black, dashed line) for codec with $b=1$, shown as a function of galaxy S/N. The black dotted line is the prediction using Eq.~(\ref{mfit}).}
\label{e1e2biasSN} 
\end{figure}

We can easily see these trends if we sort the galaxies into five wide ellipticity bins and look at the mean shifts ($e_i^{{\rm f}}-e_i^{{\rm o}}$, where ``o" denotes original images and ``f" denotes codec images), as shown in Figure~\ref{e1e2shifts}. We also find that this multiplicative bias depends on galaxy S/N, as displayed in Figure~\ref{e1e2biasSN} for the $b=1$ case, once again fitting the ellipticities from the codec images to those from the original images. This dependence is qualitatively consistent with what we find from the noise-equalization procedure.

\begin{figure}
\plotone{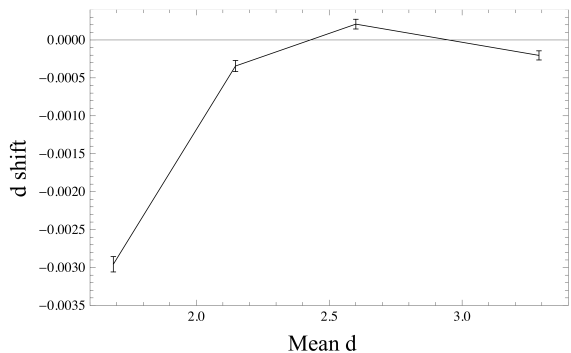}
\caption[]{Shifts in measured galaxy size from codec with $b=1$ as a function of the mean size, when the galaxies are sorted into four wide bins.}
\label{dshifts} 
\end{figure}

We can perform a similar analysis to find how our codec procedure affects the measured sizes of galaxies, given by Eq.~(\ref{size}). Let $d^{{\rm o}}$ correspond to galaxy sizes as measured in the original images and $d^{{\rm f}}$ correspond to what we measure from the codec images.  We then fit
\be
d^{{\rm f}}-d^{{\rm o}}=m_d\left(\frac{d^{{\rm f}}+d^{{\rm o}}}{2}\right)+c_d
\label{d}
\ee
to find $c_d=-0.00014 \pm 0.00005$ and $m_d=0.00006 \pm 0.00001$, for $b=1$. From Figure~\ref{dshifts}, where we bin the galaxies by size, we find that these errors come from the smallest galaxies, as is also the case with noise-equalization alone.

\section{Discussion and Recommendations}

We have studied some of the effects of applying a square-root lossy compression algorithm to images intended for weak lensing, taking the conservative approach wherein we do not make any attempt to correct for said effects.  As such, the errors found above are upper limits on what should be expected in a realistic situation, and even so we find that they are small when compared the errors resulting from the myriad of other issues that precision weak lensing surveys must confront, such as compensating for small variations in S/N.  We found no change to the sky background to within one part in $10^{6}$, a negligible increase in the shape noise, and an added digitization noise which induces a multiplicative bias on measured galaxy ellipticities and sizes.  Comparing these effects to what would happen just from adding the equivalent amount of noise, we found that the codec process combined with our shape measurement scheme leads to an excess multiplicative bias on ellipticities at the $-4\times 10^{-4}$ level for compression to $2.4$ bits per pixel.  A more sensitive test would require calibration or improvement of the shape biases in the measurement scheme. All of these results are for our fiducial WFIRST-like images, produced using our shapelets-based pipeline and analyzed with RRG.

Our study has implications for future space-based weak lensing missions such as WFIRST or Euclid. Clearly some compression is possible with a negligible loss in statistical power.  This does induce possible multiplicative shape measurement biases, but they are below the maximum level allowed as described in \citet{AR} and possibly related to limitations in the shape measurement algorithm.  Moreover, these biases can certainly be lowered by calibration with some subset of images which are not compressed and we recommend that onboard image compression be an option for future missions to allow uncompressed calibration data.  What we have demonstrated here is a method for testing the bias induced in a specific weak lensing imaging survey by a specified level of image compression.  We leave to future work the calculation of the allowable compression for any specific survey design.

There are a few things to note about our method.  For one, the pipeline used to manufacture our simulated images is somewhat simplified, as the PSF lacks sharp features like diffraction spikes and it is constant and uniform across the field.  We do not add a realistic shear signal to the galaxy images, but the typical cosmological signal is an order of magnitude less than the intrinsic shape noise of field galaxies.  We assume a constant PSF across the field and do not perform PSF deconvolution since we are interested in shape changes rather than very accurate absolute shape measurements.  As such we believe that our shapelets-based package is sufficiently realistic so that the compression-induced effects of extra shape noise and ellipticity bias should be the same in real data.  The addition of cosmic rays may reduce the compression ratio achievable with the compression level discussed here (Bernstein et al.~2010), but to what extent is extremely mission-specific and is still very uncertain for L2. One further caveat is that we did not explore other survey options, and it may be that the compression effects are sensitive to some of these options.  We further have not attempted to show that detector non-linearities could be successfully removed from codec images in the same way they could be removed from images that had not been compressed.  Finally, our weak lensing pipeline is somewhat simplified, in that we used detections and sky measurement from the original images, and we also only used one shape measurement algorithm.

Nevertheless we have shown what generically happens to weak lensing data when it has been compressed using this square-root algorithm for a simulated survey that serves as a good example of what will likely be expected in next-generation space-based weak lensing missions.  Once the actual survey strategy is determined, we will do more specific simulations to pinpoint exactly how much compression would be acceptable for a given cosmological parameter error threshold.  There is also the possibility that this bias could be calibrated if it could be accurately enough characterized.  The benefits of such calibration and potential strategies for its implementation are left to future work.

\begin{acknowledgments}
This work was supported in part by the Kavli Institute for Cosmological Physics at the University of Chicago through grants NSF PHY-0114422 and NSF PHY-0551142 and an endowment from the Kavli Foundation and its founder Fred Kavli. This work was also carried out in part at the Jet Propulsion Laboratory, California Institute of Technology, under a contract with NASA, and funded by JPL's Research and Technology Development Funds. CS and DJ acknowledge support from the Fermi Research Alliance, LLC under Contract No. DE-AC02-07CH11359 with the United States Department of Energy. GMB acknowledges support from grant AST-0607667 from the NSF and DOE grant DE-FG02-95ER40893. RM acknowledges support from an STFC Advanced Fellowship and from ERC grant MIRG-CT-208994.
\end{acknowledgments}

\end{document}